\documentclass[11pt,twoside]{article}

\usepackage{asp2006}
\usepackage{epsf}
\usepackage{psfig}
\usepackage{lscape}
\usepackage{amssymb}

\def\cm3{\hbox{cm$^{-3}$}}

\def\one{\,{\sc i}}

\newcommand\hst{{\it HST}}
\newcommand\ie{i.\,e}

\newcommand\eg{e.\,g.}

\newcommand\spit{\textit{Spitzer}}

\newcommand\bvi{\textit{BVI}}

\begin{document}
\title{Star clusters as tracers of galaxy evolution}
\author{Iraklis S. Konstantopoulos}
\affil{The Pennsylvania State University}

\begin{abstract}
Star clusters represent the most common `mode' of star formation. They are found in all types of environments, cascading down from galaxy groups and merging pairs through starbursts to normal galaxies and dwarves and even isolated regions in extragalactic space. As they maintain a link to the overall star formation in a system, they can be used as tracers of the star formation history of environments located at distances prohibitive to the study of individual stars. This makes them ideally suited to the study of mergers and interactions in galaxy pairs and groups. In this work we present observations of the star cluster populations in the local starburst galaxy M82, post-interaction spiral NGC~6872, the ``Antenn\ae '' merging pair and two compact groups, ``Stephan's Quintet'' and HCG~7. In each case, we extract information on the clusters and their hosts using mainly \hst\ photometry and Gemini spectroscopy. 
\end{abstract}

\section{Introduction}\label{sec:intro}
Star cluster formation is not limited to the `classical' environments of galactic spiral arms and bright bulges. Over the past two decades, observations have found them in all types of quiescent lieus, a discovery that contributed to their characterisation as the single `mode' of star formation. As such, they are quickly becoming an alternative to stars as tracers of star formation in the environments in which they are formed and evolve. A great advantage presented by star clusters is that they shine bright even at extreme distances, beyond the boundaries where individual stars perish below our detection limits. 

The current state of star cluster research allows for many assumptions to be made safely, based on the large volume of existing information. One of the first discoveries was that of a power-law luminosity function, a distinguishing factor between young and globular clusters. A good understanding of star cluster disruption has helped to decipher the observed age distributions and allow for the hypothesis to present itself that young massive clusters (YMC) are equivalent to the methuselaic globulars -- once thought to be a unique class in the history of the universe. And as time goes on, more of the fundamental properties of cluster populations are being revised, revisited and updated -- \eg\ it has been suggested that the shape of the initial mass function (CIMF) is well described by a Schechter function \citep{gieles09a,larsen09}. This results in a luminosity function (LF) with a luminosity-dependent logarithmic slope, which steepens towards brighter luminosities \citep[see the contribution by Gieles;][]{gieles09b}.

The described knowledge of cluster populations allows for the extension of their use to far away environments, where only the brightest part of the LF is observable. Understanding the underlying statistics is essential to drawing results in such environments, a comprehension that we are very close to possessing fully. Following this path, we present our studies of young star clusters in a number of interacting systems and test the extent to which our high-quality imaging/photometry and spectroscopy can reach in terms of studying these systems through their cluster populations.

\section{M82: 200~Myr of formation and evolution in the original starburst}\label{sec:m82}
In the context of this study, M82 provides a nearby test-case. This is a starburst galaxy that has been producing clusters at a high rate over the past $\sim220$~Myr, the time of the triple interaction in which M81 and NGC~3077 took part. The interaction time-scale was derived by \cite{yun94}, through the modelling of the H\one\ gas in the M81 group (based on high resolution VLA data). In our recent studies of the galaxy \citep{smith07,isk08,isk09a}, we have provided the first observational evidence towards the age of this interaction. By accurately age-dating each of 49 clusters in our sample, we derive the cluster age distribution that we relate to the starburst history of the galaxy: after a period of low SF activity, we find a burst at $\sim250$~Myr; the burst spreads across the disk over the following $\sim50$~Myr, reaching the epoch of maximum star formation at $\tau\sim150$~Myr; this is then followed by a gradual decline in cluster formation rate (CFR), until the only actively star-forming region is the nucleus. This is very well matched to the theoretical scenario offered by \citet{forster03}. 

\begin{figure}[t]
	\plottwo{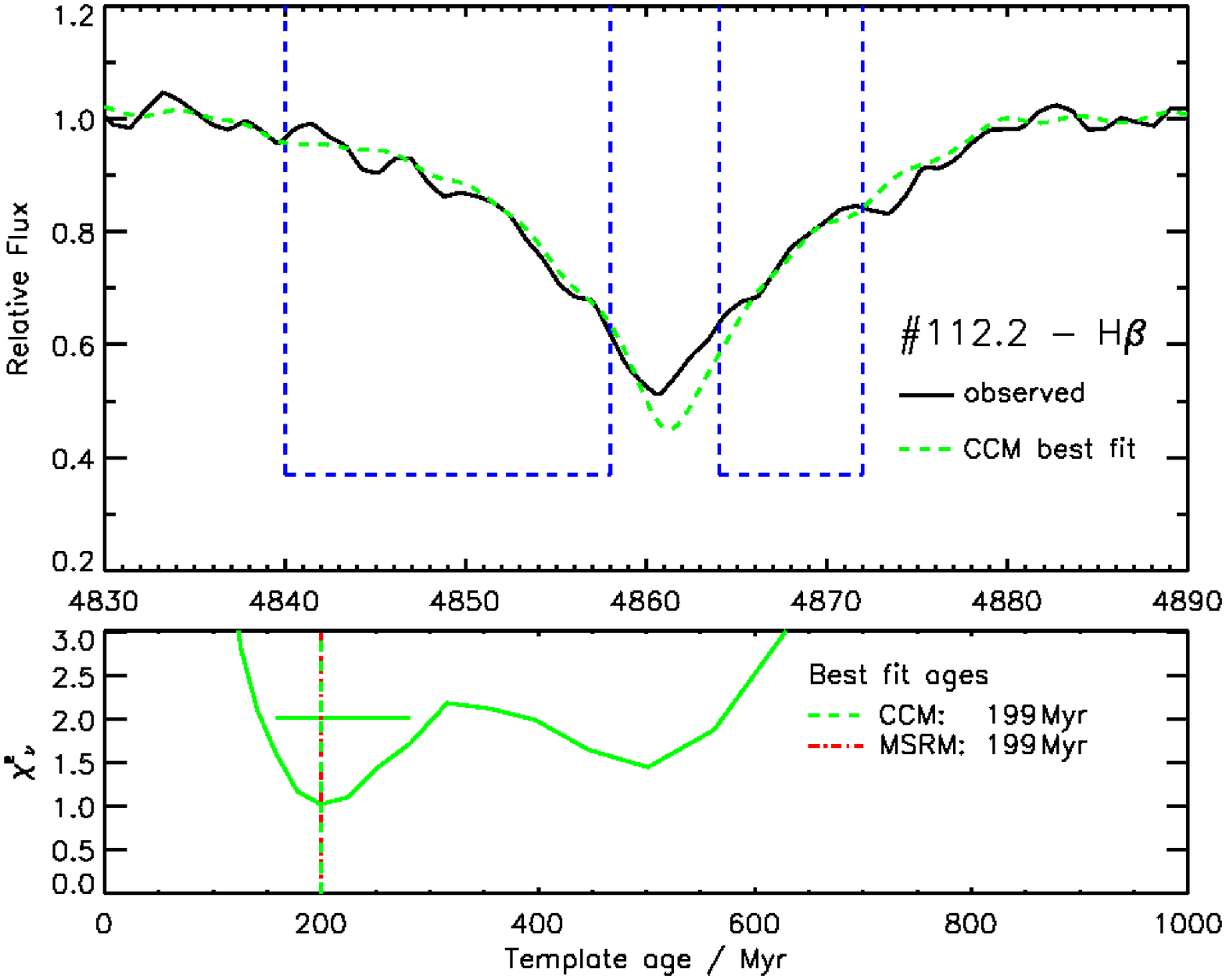}{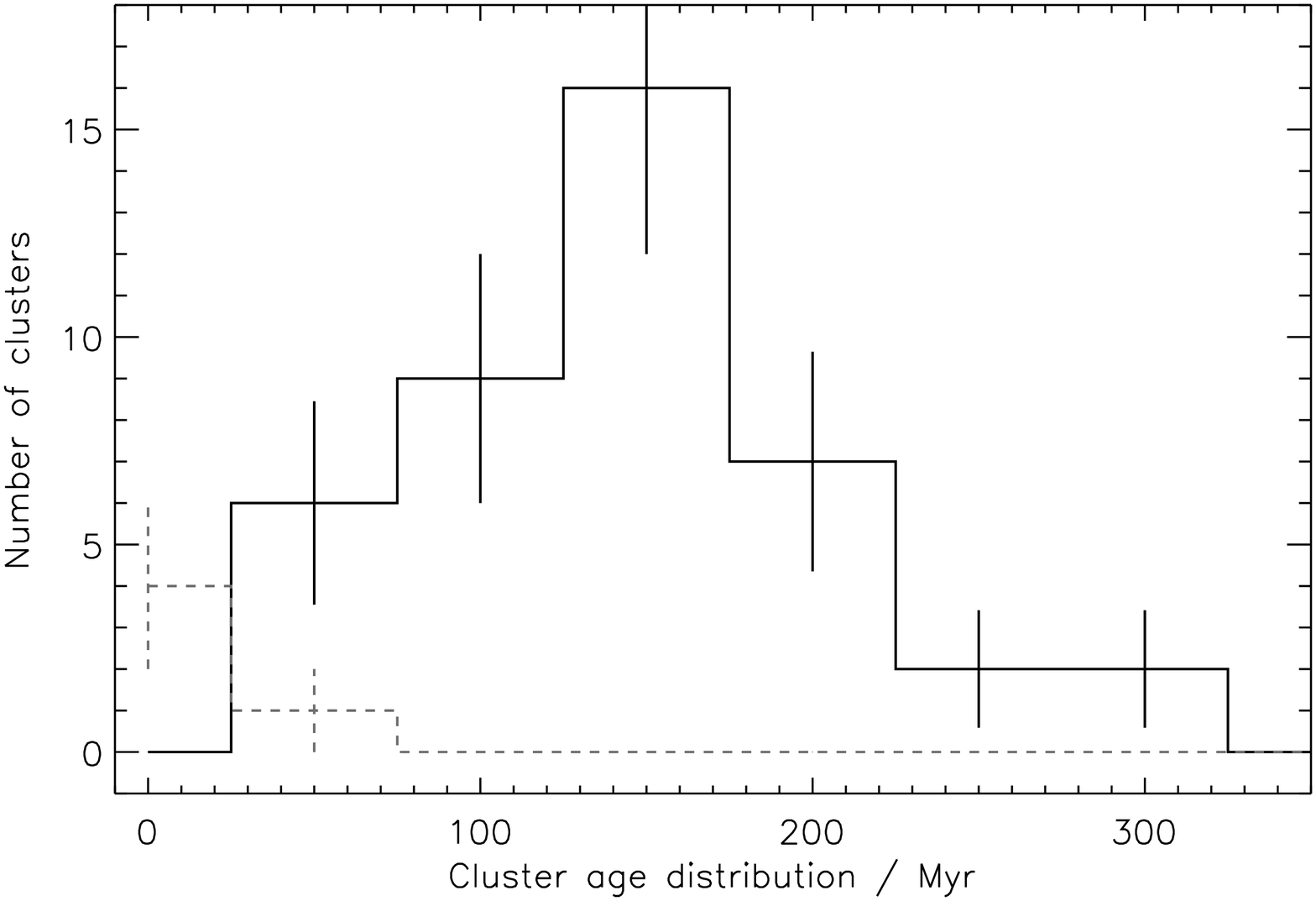}
	\caption{The {\bf left} panel shows the direct age fitting process employed for clusters in M82. Since the metallicity of the M82 ISM is known, we fit with age as the sole variable. After obtaining a reliable age for each cluster, we plot the age distribution ({\bf right}). This shows the cluster formation history and, as star clusters express the preferred mode of star formation, the starburst history of M82. From this plot it follows that, after a period of low activity, the galaxy undergoes a burst of star formation at $\sim250$~Myr. A further $\sim50$~Myr later, the starburst has spread across the disk. After reaching this epoch of maximum star formation, the activity declines steadily, until at a few Myr in the past, cluster formation is confined to the nuclear region (dashed line). 
}\label{fig:m82-age}

\end{figure}

\section{NGC~6872: beads on a string}\label{sec:n6872}
NGC~6872 is one half of an interacting pair of galaxies. While its partner, IC~4970, appears to have maintained a fairly normal morphology, NGC~6872 has developed imposing, long tidal features, on which one can easily spot cluster formation as `beads on a string'. Most interestingly, this new wave of star formation seems to only be affecting the tidal tails and not the main body of the galaxy. 

\begin{figure}[h]
	\plotone{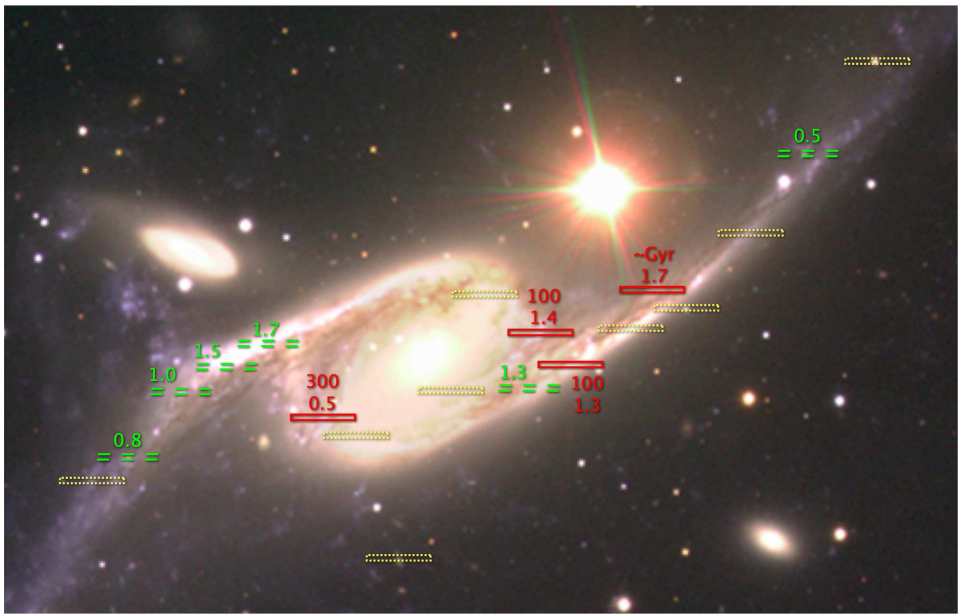}	
	\caption{Optical imaging of NGC~6872 (VLT) with Gemini MOS slits over-plotted: red (solid) slits represent absorption line objects, \ie\ post gas-expulsion (age~$\gtrsim20$~Myr) and we quote the age in Myr and the metallicity ($Z$) for each cluster, in solar units; green (dashed) slits denote spectra with emission line features, that are most likely shrouded in their natal gas (this can be inferred by the spectral signatures of emission and absorption), with $Z$ quoted. As this work is in progress, yellow (dotted) slits are yet to be studied. 
	}\label{fig:n6872}
	\end{figure}

In this galaxy, we studied a handful of clusters in the tidal tails following the methodology described in \cite{trancho07}. We found them to belong to a generation of star formation commencing no longer than $\sim100$~Myr ago. A very interesting observation pertains to the metallicity of these clusters, that displays a clear outwardly decreasing gradient -- from super-solar nearest to the body, to sub-solar in the mid-extent of the tail. More information is provided in Fig.~\ref{fig:n6872}. 

\section{The Antenn\ae : orderly formation in a chaotic environment}\label{sec:ant}
The Antenn\ae\ are one of the most studied environments in the local universe. They provide an exquisite laboratory for the study of star cluster formation and merger evolution. Here we present the results of a spectroscopic study of 15 clusters in the main body and tail of the galaxies. We obtain ages and metallicities through the Trancho method. The most striking feature of this population is that 13 out of the 16 clusters seem to be co-rotating with one of the progenitor galaxy disks. This suggests that, even well into the evolution of this merger, star formation is still proceeding in an orderly fashion. We are therefore witnessing the first `randomisations' of cluster orbits, well before they are kicked into the outskirts. These results were extracted from \cite{bastian09}; we refer the reader to this article for figures and further information.

\section{Stephan's Quintet: multiple interactions and tidal tail formation}\label{sec:s5}
Over the preceding sections we demonstrated the use of star clusters as tracers of galaxy interactions. We will now apply the same techniques to far away galaxy groups, where clusters are invaluable as tracers of star formation. 

\begin{figure}[h]

	\plotone{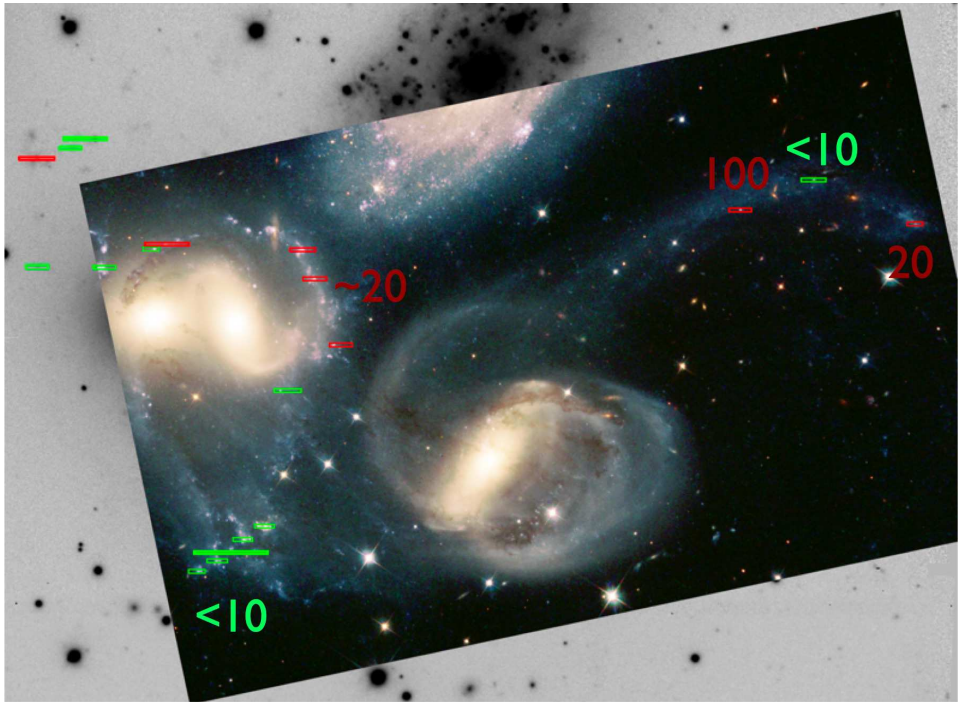}
	\caption{The multiple interaction in S5, as captured by \hst~(south is to the right; GMOS imaging in the background). We used spectroscopy to confirm previously derived photometric ages in different regions (colour-coding follows Fig.~\ref{fig:n6872}). Again, we find ongoing formation since perigalacticon. Interestingly, the clumps on the west side of the NGC~7318 galaxy pair (A, B in a left to right arrangement) are apparently being stripped of their natal gas, possibly due to an external tidal stripping mechanism. The age, size and mass of the object at the edge of the TT suggests that it may be a tidal dwarf galaxy in formation. 
	}\label{fig:s5}

	\end{figure}

The first group we visit is ``Stephan's Quintet'', a merging/interacting system of five members situated at a distance of $\sim100$~Mpc (Fig.~\ref{fig:s5}). At the great distance to this group, we can resolve the high end of the cluster luminosity function (CLF), which translates to high-mass clusters, as we expect most detectable clusters to be young ($\tau<1$~Gyr). We have obtained spectra of clusters in the tidal tail (TT) south of NGC~7319, as well as the northern starburst region (NSR) and the overlap region (OR) between NGC~7318/9. Spectral ages (again obtained through the Trancho method) agree with the photometric age-dating by \cite{gallagher01}: between $10-100$~Myr in the TT, $\lesssim10$~Myr in the NSR and $\lesssim20$~Myr in the OR (slightly older than previously predicted). In the OR we find some clusters that are best described by ages of $<10$~Myr, but are already devoid of their natal gas cloud. Their location, however, is extremely bright in X-rays, indicating the presence of a strong shock front. One possibility is therefore that they have been stripped of their natal gas entirely (or to a great extent) by an external process. The fact that some of their neighbours have survived to $\sim20$~Myr, coupled with their high masses, might thus offer some indication that massive clusters are more stable against disruption. 

A bonus offered by the use of spectra is the radial velocity (RV) information contained. We used the cluster RVs to relate them to one of the galaxies, thus establishing that the tail belongs to galaxy NGC~7318A (see Fig.~\ref{fig:s5}).

\section{HCG~7: the path toward a dry merger}
As shown in the previous section, the study of compact groups can be greatly enhanced by an understanding of their cluster populations. We used \hst\ and \spit\ imaging and VLA H\one\ maps of HCG~7, a group of three late-type and one early-type galaxy,  to derive its interaction record. Interestingly, we find that record to be blank, with no interaction history between the four members of this group. This is testified by the distribution of H\one\ gas that is contained entirely within the individual galaxies. Interestingly, the high star formation rates in the individual spirals, as derived by both the cluster formation rate and their FIR luminosities, implies that the galaxies will deplete their reservoir of Hydrogen gas before they start merging. This would produce a massive elliptical galaxy with no recent SFH and no X-ray halo. We therefore speculate that HCG~7 is the precursor to a `dry merger'. 

	\begin{figure}[h]

	\plotone{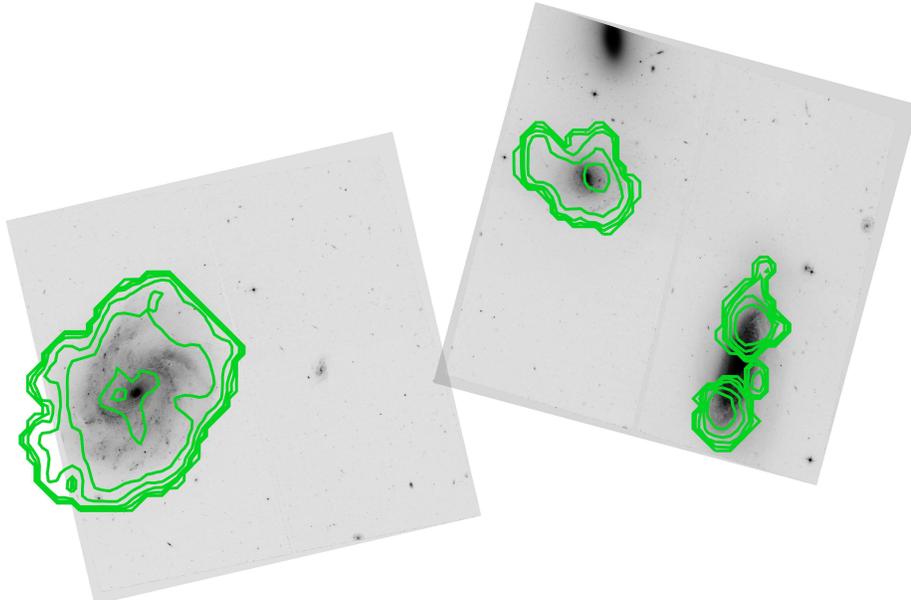}
	\caption{\hst-ACS image of HCG~7 (\bvi\ composite), with H\one\ contours over-plotted. The absence of gas from the intra-group medium is indicative of a blank interaction history between the galaxies in this group. The high star formation rates of the early-type members suggest that the gas will be depleted long before the galaxies begin to interact. Thus, HCG~7  is quite plausibly heading toward a `dry merger', where the end state is a massive elliptical galaxy with no X-ray halo. 
	}\label{fig:hcg7}

	\end{figure}

\section{Summary}\label{sec:summary}
We have presented a number of studies of interacting/merging galaxy pairs and groups, based on their populations of star clusters. We have used cluster spectra and colours to directly draw information on the interaction timescale, the metallicity of the environment and galaxy dynamics. We have used this information to derive the current dynamical state of these systems and lend support to scenaria for the evolution of isolated starbursts and compact groups as a whole. In all, we have demonstrated the diagnostic strength of star clusters for describing and characterising various classes of interacting systems.

\acknowledgements 
I would like to acknowledge the financial support and thank the staff of Gemini South for their hospitality during the undertaking of parts of the presented work. Support for this work was also provided by NASA through grant number HST-GO-10787.15-A from the Space Telescope Science Institute which is operated by AURA, Inc., under NASA contract NAS 5-26555, and the National Science and Engineering Council of Canada (SCG).

\end{document}